\documentclass[print, twocolumn,showpacs]{revtex4}
\usepackage{amsmath}
\usepackage{txfonts}
\usepackage{graphicx}
\usepackage{bm}
\usepackage{epsfig}
\usepackage{setspace}
\begin{document}
\draft
\bibliographystyle{revtex4}
\titlepage

\title{Dynamics of quantum correlation for a central two-qubit coupled to an isotropic Lipkin-Meshkov-Glick bath}
\author{Yi-Ying Yan}\author{Li-Guo Qin}\author{Li-Jun Tian\footnote{Email: tianlijun@staff.shu.edu.cn.}}
\affiliation {Department of Physics, Shanghai University, Shanghai,
200444, China}

\begin{abstract}
We investigate the behavior of quantum correlation, measured by quantum
discord and entanglement, for two central spin qubits coupled to an
isotropic Lipkin-Meshkov-Glick bath. We find that when the bath falls
into different phases, both measurements of quantum correlation follow
distinct paths. When the bath is in the symmetry-broken phase, as
tuning the interaction between two-qubit and the bath being close
to the interaction of bath spins, quantum discord and entanglement
remain stable values with slight fluctuations. However, in the bath
with the symmetric phase, quantum discord and entanglement always
vary periodically with sharp fluctuations. Furthermore, the fluctuations
can be enhanced by increasing the coupling strength between qubits
and the bath. The critical point of quantum phase transition of the
bath can be revealed clearly by the distinct behaviors of quantum
correlation in the different phases of the bath. Besides, it is observed
that quantum discord is significantly enhanced during the evolution
while entanglement periodically vanishes.
\end{abstract}
\maketitle

\section{Introduction}
Nature utilizes correlation to bridge different things. Entanglement,
a kind of quantum correlation, has been taken to be fundamental resources
for quantum information and communication~\cite{Nielsen2,Horodecki1}. Extensive studies
have shown that quantum entanglement is absolutely indispensable in
such issues like superdense coding~\cite{Bennett2}, quantum teleportation~\cite{Bennett3},
quantum cryptography~\cite{Ekert1}, etc. However, it was found that there
are some types of nonclassical correlation other than entanglement
that also provide some advantages for the quantum systems over their
classical counterparts~\cite{Zurek1,Vedral2}. Such a non-entanglement quantum correlation
can be captured by quantum discord~\cite{Zurek1}, which is defined as the
difference between quantum mutual information and classical correlation.
Investigations have shown that quantum discord is responsible for
the computational efficiency of deterministic quantum computation
with one pure qubit (DQC1)~\cite{Knill1,Datta1,Lanyon1}. One feature of this intrinsically quantum
correlation lies on that fact that quantum discord may have nonzero
value for certain separable states, while entanglement vanishes~\cite{Zurek1,Luo1}.
In general, for pure states quantum discord and entanglement entropy
are equivalent, otherwise they are different.

The difference between quantum discord and entanglement as figures
of merit for characterizing quantum correlation is expected to be
inherited and manifested in the phenomena of quantum phase transitions
(QPTs)~\cite{Dillenschneider1,yaoc1,Maziero1,Werlang3}. On one hand, since quantum discord grows towards the
critical points in the context of QPTs, it work effectively as well
as entanglement to signal QPTs. On the other hand, quantum
discord may be more advantageous than entanglement at some conditions. For instance, pairwise quantum discord in infinite $XY$ chain may be able to signal QPT for distances between sites where pairwise entanglement vanishes~\cite{Maziero1}. For $XXZ$ spin chain at finite temperature, quantum discord spotlights the critical point with respect to QPT while both entanglement and thermaldynamics quantity fails~\cite{Werlang3}. More recently, it was claimed that quantum discord is the best critical point estimator when one deals with finite-temperature systems~\cite{Werlang4}.

For realistic quantum systems, the loss of coherence is inevitable due to the interaction with the environment~\cite{Breuer1}. Recently, there have been increasing researches on quantum discord dynamics under the effect of the environment~\cite{Werlang5,Mazzola1,Vasile1,Wang3,Fanchini1}. In the Markovian environment, it has been shown that in all cases when entanglement suddenly disappears, quantum discord vanishes only in the asymptotic limit~\cite{Werlang5}. Thus, it was argued that quantum algorithms based on quantum discord can be more robust than those based on entanglement. In the non-Markovian environment, the decline of quantum discord can be altered.~\cite{Vasile1,Wang3,Fanchini1}.

Spin environment has been obtained much attentions~\cite{Hutton1, Cucchietti1, Rossini1, Quan2, Yuan1, Cheng1,Cheng2}. Quantum coherence of a central two-qubit is rapidly destroyed as the environment being close to the critical point of its QPT~\cite{Quan2,Yuan1}. Liu \textit{et al.} have considered the effects of $XY$ spin chain environment on quantum discord of a central two-qubit and showed that quantum discord can become minimized close to the critical point of a QPT of $XY$ chain~\cite{Liu1}. Previously, Lipkin-Meshkov-Glick (LMG) model~\cite{LMG1}, originally introduced in nuclear physics, has been proved to have a QPT~\cite{Vidal1,Dusuel1,Dusuel2}. Purity of single central qubit coupled to an isotropic LMG bath witnesses the QPT critical point of the system~\cite{Quan1}. Therefore it is intuitive to ask what would happen for quantum correlation when a central two-qubit couples to such an isotropic LMG bath.

In this paper, we will analyze the dynamics of quantum correlation of a central two-qubit, which interacts with an isotropic LMG bath. It is of interest to reveal the influence of the QPT on quantum correlation and the distinct behaviors of quantum correlation when bath is in two different phases. The paper is organized as follow. In Sec.~\ref{Dyn}, we introduce our physical model and obtain the reduced density matrix of
central two-qubit for two different phases. In Sec.~\ref{QC}, we calculate the quantum discord and entanglement, and present the main results. The last section is devoted to conclusions.
\section{Dynamics of a central two-qubit coupled to an isotropic LMG bath}\label{Dyn}
We consider two noninteracting central spin qubits coupled to an
isotropic LMG bath. The total Hamiltonian can be expressed as
\begin{equation}
H  =  H_{B}+H_{I}+H_{S},
\end{equation}
with
\begin{eqnarray}
H_{B} & = & -\frac{\lambda}{N}\sum_{i<j}^{N}(\sigma_{i}^{x}\sigma_{j}^{x}+\sigma_{i}^{y}\sigma_{j}^{y})-\sum_{i=1}^{N}\sigma_{i}^{z},\nonumber\\
H_{I} & = & -\frac{\lambda^{\prime}}{N}\sum_{i=1}^{N}[\sigma_{i}^{x}(\sigma_{a}^{x}+\sigma_{b}^{x})+\sigma_{i}^{y}(\sigma_{a}^{y}+\sigma_{b}^{y})],\nonumber\\
H_{S} & = & -\sigma_{a}^{z}-\sigma_{b}^{z},
\end{eqnarray}
where $H_{B}$ represents the Hamiltonian of the isotropic LMG bath, $H_{I}$ denotes the interaction Hamiltonian between a central two-qubit and the bath, and $H_{S}$ describes the central two-qubit. Here $\sigma^{\alpha}_{i}, (\alpha=x,y,z, i=1,2,\cdots,N)$ are the Pauli matrices of the $i-$th spin, $\sigma^{\alpha}_{a(b)}$ are the Pauli matrices of the central qubit $a(b)$. $N$ is the spin numbers of the bath, $\lambda$ is the coupling strength of the spins in the bath, $\lambda^{\prime}$ the coupling strength between the central two-qubit and the bath. In the Dicke representation \cite{Vidal1,Dusuel1}, the above Hamiltonian can be rewritten as

\begin{equation}
H=-\frac{\lambda}{N}(S_{N}^{+}S_{N}^{-}+S_{N}^{-}S_{N}^{+}-N)-2S_{N}^{z}-\frac{2\lambda^{\prime}}{N}(S_{+}S_{N}^{-}+S_{-}S_{N}^{+})-2S_{z},
\end{equation}
where $S_{N}^{\pm}=S_{N}^{x}\pm iS_{N}^{y}$ with $S_{N}^{\alpha}=\frac{1}{2}\sum_{i=1}^{N}\sigma_{i}^{\alpha}$, $S_{\pm}=\frac{1}{2}(\sigma_{A}^{x}+\sigma_{B}^{x})\pm\frac{i}{2}(\sigma_{A}^{y}+\sigma_{B}^{y})$ and $S_{z}=\frac{1}{2}(\sigma_{A}^{z}+\sigma_{B}^{z})$. The groundstate of isotropic LMG model lies in the subspace spanned by the Dicke states $\{|N/2,M\rangle,M=-N/2,\cdots,N/2\}$ \cite{Vidal1,Dusuel1}. It is easy to show that the energy corresponding to the Dicke states is $\lambda$ dependent, so the groundstate $|G\rangle$ is~\cite{Dusuel2,Latorre1},
\begin{equation}
|G\rangle=
\begin{cases}
\bigg|\frac{N}{2},\frac{N}{2}\bigg\rangle & \text{$(0<\lambda<1)$}, \\
\bigg|\frac{N}{2},I(\lambda)\bigg\rangle  & \text{$(\lambda>1)$},
\end{cases}
\end{equation}
where $I(\lambda)$ denotes the integer nearest to $N/(2\lambda)$. It implies that there is a QPT at critical point $\lambda=1$. For the case of $\lambda>1$, the LMG bath is in a symmetric phase. For the case of $0<\lambda<1$, the LMG bath is in a symmetry-broken phase \cite{Dusuel2,Latorre1,Barthel1}. We below would study the dynamics of quantum correlation for the bath in each phase.
We simply denote triplets and singlet of two central spins by $|1,1\rangle=|\uparrow\uparrow\rangle, |1,0\rangle=\frac{1}{\sqrt{2}}(|\uparrow\downarrow\rangle+|\downarrow\uparrow\rangle)$, $|1,-1\rangle=|\downarrow\downarrow\rangle$, and $|0,0\rangle=\frac{1}{\sqrt{2}}(|\uparrow\downarrow\rangle-|\downarrow\uparrow\rangle)$, respectively.

In the invariant subspace $H_{M}$ of $H$, spanned by the ordered basis $\{|N/2,M\rangle\otimes|1,1\rangle, |N/2,M+1\rangle\otimes|1,0\rangle, |N/2,M+2\rangle\otimes|1,-1\rangle\}$, the total Hamiltonian can be expressed as a quasidiagonal matrix with the blocks~\cite{Quan1}

\begin{equation}\label{HM}
H_{M}=\left(\begin{array}{ccc}
\alpha & \xi & 0\\
\xi & \beta & \kappa\\
0 & \kappa & \gamma
\end{array}\right),
\end{equation}
with
\begin{eqnarray}
\alpha & = & -\frac{\lambda}{2N}(N^{2}-4M^{2})-2(M+1),\nonumber\\
\beta & = & -\frac{\lambda}{2N}[N^{2}-4(M+1)^{2}]-2(M+1),\nonumber\\
\gamma & = & -\frac{\lambda}{2N}[N^{2}-4(M+2)^{2})]-2(M+1),\nonumber\\
\xi & = & -\frac{\lambda^{\prime}}{N}\sqrt{2N(N+2)-8M(M+1)},\nonumber\\
\kappa & = & -\frac{\lambda^{\prime}}{N}\sqrt{2N(N+2)-8(M+1)(M+2)}.
\end{eqnarray}

The eigenvector corresponding to the eigenvalue $E_{j}(j=1,2,3)$ of $H_{M}$ is directly obtained as
\begin{equation}
|\Psi_{j}\rangle=p_{j}\bigg|\frac{N}{2},M\bigg\rangle\otimes|1,1\rangle+q_{j}\bigg|\frac{N}{2},M+1\bigg\rangle\otimes|1,0\rangle+r_{j}\bigg|\frac{N}{2},M+2\bigg\rangle\otimes|1,-1\rangle,
\end{equation}
with
\begin{eqnarray}
p_{j} & = & \frac{\xi(E_{j}-\gamma)}{\sqrt{(E_{j}-\alpha)^{2}(E_{j}-\gamma)^{2}+(E_{j}-\gamma)^{2}\xi^{2}+(E_{j}-\alpha)^{2}\kappa^{2}}},\nonumber\\
q_{j} & = & \frac{(E_{j}-\alpha)(E_{j}-\gamma)}{\sqrt{(E_{j}-\alpha)^{2}(E_{j}-\gamma)^{2}+(E_{j}-\gamma)^{2}\xi^{2}+(E_{j}-\alpha)^{2}\kappa^{2}}},\nonumber\\
r_{j} & = & \frac{\kappa(E_{j}-\alpha)}{\sqrt{(E_{j}-\alpha)^{2}(E_{j}-\gamma)^{2}+(E_{j}-\gamma)^{2}\xi^{2}+(E_{j}-\alpha)^{2}\kappa^{2}}}.
\end{eqnarray}
For simplicity, we consider the cases of $H_{M}$ with three distinct eigenvalues. Thus, $\{|\Psi_{j}\rangle\}$ is a set of orthonormal basis of $H_{M}$. Then $H_{M}$ is straightforward diagonalized as $T^{-1}H_{M}T=$diag$\{E_{1},E_{2},E_{3}\}$ with the matrix
\begin{equation}
T = \left(\begin{array}{ccc}
p_{1} & p_{2} & p_{3}\\
q_{1} & q_{2} & q_{3}\\
r_{1} & r_{2} & r_{3}
\end{array}\right).
\end{equation}
The dynamics evolution operator $U(t)=\exp(-iHt)$ in the subspace $H_{M}$ can be obtained as~\cite{Quan1}
\begin{eqnarray}\label{UT1}
U_{M}(t) & = & T\left(\begin{array}{ccc}
e^{-iE_{1}t}\\
 & e^{-iE_{2}t}\\
 &  & e^{-iE_{3}t}
\end{array}\right)T^{-1}\nonumber\\
 & = & \left(\begin{array}{ccc}
\sum_{j=1}^{3}p_{j}^{2}e^{-iE_{j}t} & \sum_{j=1}^{3}p_{j}q_{j}e^{-iE_{j}t} & \sum_{j=1}^{3}p_{j}r_{j}e^{-iE_{j}t}\\
\sum_{j=1}^{3}p_{j}q_{j}e^{-iE_{j}t} & \sum_{j=1}^{3}q_{j}^{2}e^{-iE_{j}t} & \sum_{j=1}^{3}q_{j}r_{j}e^{-iE_{j}t}\\
\sum_{j=1}^{3}p_{j}r_{j}e^{-iE_{j}t} & \sum_{j=1}^{3}q_{j}r_{j}e^{-iE_{j}t} & \sum_{j=1}^{3}r_{j}^{2}e^{-iE_{j}t}
\end{array}\right)\nonumber\\
 & \equiv & \left(\begin{array}{ccc}
U_{11} & U_{12} & U_{13}\\
U_{12} & U_{22} & U_{23}\\
U_{13} & U_{23} & U_{33}
\end{array}\right).
\end{eqnarray}
It is worth noting that the Eq.~(\ref{UT1}) is valid for the cases $-N/2\leq M<N/2-1$, as $\{|N/2,M\rangle\otimes|1,1\rangle,|N/2,M+1\rangle\otimes|1,0\rangle,|N/2,M+2\rangle\otimes|1,-1\rangle\}$ is a three-dimensional invariant subspace for these cases. When $M=N/2-1$, $\{|N/2,N/2-1\rangle\otimes|1,1\rangle,|N/2,N/2\rangle\otimes|1,0\rangle\}$ forms a two-dimensional invariant subspace $H_{N/2-1}$, and the dynamics evolution operator in this subspace can be calculated by the same procedure by setting $\kappa=0$ and $\gamma=0$ in Eq.~(\ref{HM}). Besides, the states $|N/2,N/2\rangle\otimes|1,1\rangle$ and $|N/2,M\rangle\otimes|0,0\rangle$ belong to one-dimensional invariant subspace, namely, they are the eigenstates of the total Hamiltonian $H$, the corresponding eigenenergies are $-(N+2)$ and $2\lambda M^{2}-\lambda N/2-2M$, respectively. Thus the dynamics evolution of these states are $\exp[i(N+2)t]$ and $\exp[-i(2\lambda M^{2}-\lambda N/2-2M)t]$.

To highlight the behavior of quantum correlation with respect to QPT, it is natural to assume the bath to be in the groundstate $|G\rangle$. Besides, we suppose that the central two-qubit is initially prepared in a class of $X-$structure state, $\rho_{AB}(0)=\frac{1}{4}(I+\sum_{\alpha}k_{\alpha}\sigma_{A}^{\alpha}\otimes\sigma_{B}^{\alpha})$. Here $I$ is the identity operator on two-qubit, and $k_{\alpha}$ $(\alpha=x,y,z)$ are real parameters to make $\rho_{AB}(0)$ a legal quantum state. The evolution state of the total system is given by $\rho_{tot}(t)=U(t)\rho_{tot}(0)U^{\dagger}(t)$, where $\rho_{tot}(0)=|G\rangle\langle G|\otimes\rho_{AB}(0)$ is the initial state of the total system.

\subsection{The LMG bath with symmetric phase}
When the bath is in symmetric phase $(\lambda>1)$, $I(\lambda)<N/2$ and the groundstate is $|G\rangle=|N/2,I(\lambda)\rangle$. One would notice that $|N/2,M\rangle\otimes|1,1\rangle$, $|N/2,M\rangle\otimes|1,0\rangle$ and $|N/2,M\rangle\otimes|1,-1\rangle$ belong to three different invariant subspaces, $H_{M}$, $H_{M-1}$, and $H_{M-2}$, respectively. By the observation, we can apply Eq.~(\ref{UT1}) to obtain the evolution state. The reduced density matrix of central two-qubit is obtained by tracing out the bath, and in the basis $\{|\uparrow\uparrow\rangle,|\downarrow\downarrow\rangle,|\uparrow\downarrow\rangle,|\downarrow\uparrow\rangle\}$, it takes the form
\begin{equation}\label{rho1}
\rho_{ab}(t)=Tr_{bath}[U(t)\rho_{tot}(0)U^{\dagger}(t)]=\left(\begin{array}{cc}
\mathcal{A} & \mathcal{Z}\\
\mathcal{Z}^{*} & \mathcal{C}
\end{array}\right)\oplus\left(\begin{array}{cc}
\mathcal{B} & \mathcal{Y}\\
\mathcal{Y} & \mathcal{B}
\end{array}\right),
\end{equation}
with
\begin{eqnarray}\label{var1}
\mathcal{A} & = & \frac{1+k_{z}}{4}(|U_{11}|^{2}+|U_{13}^{\prime\prime}|^{2})+\frac{1+k_{x}+k_{y}-k_{z}}{4}|U_{12}^{\prime}|^{2},\nonumber\\
\mathcal{B} & = & \frac{1+k_{z}}{8}(|U_{12}|^{2}+|U_{23}^{\prime\prime}|^{2})+\frac{1+k_{x}+k_{y}-k_{z}}{8}|U_{22}^{\prime}|^{2}\nonumber\\
  &   &+\frac{1-k_{x}-k_{y}-k_{z}}{8},\nonumber\\
\mathcal{C} & = & \frac{1+k_{z}}{4}\bigg[\theta\bigg(\frac{N}{2}-I(\lambda)-1\bigg)|U_{13}|^{2}+|U_{33}^{\prime\prime}|^{2}\bigg]\nonumber\\
 &  &+\frac{1+k_{x}+k_{y}-k_{z}}{4}|U_{23}^{\prime}|^{2},\nonumber\\
\mathcal{Y} & = & \frac{1+k_{z}}{8}(|U_{12}|^{2}+|U_{23}^{\prime\prime}|^{2})+\frac{1+k_{x}+k_{y}-k_{z}}{8}|U_{22}^{\prime}|^{2}\nonumber\\
 &  &-\frac{1-k_{x}-k_{y}-k_{z}}{8},\nonumber\\
\mathcal{Z} & = & \frac{k_{x}-k_{y}}{4}U_{11}U_{33}^{\prime\prime*},
\end{eqnarray}
where $U_{ij}$,$U^{\prime}_{ij}$, and $U^{\prime\prime}_{ij}$ $(i,j=1,2,3)$ denote the matrices elements of the dynamics evolution operator in the invariant subspaces $H_{I(\lambda)}$, $H_{I(\lambda)-1}$, and $H_{I(\lambda)-2}$, respectively. $\theta(x)$ is the Heaviside step function, which equals to $0$ for $x<0$ and $1$ for $x\geqslant0$.

\subsection{The LMG bath with symmetry-broken phase}
When the bath is in symmetry-broken phase $(0<\lambda<1)$, the groundstate is $|G\rangle=|N/2,N/2\rangle$. As mentioned above, the dynamics evolution of the states $|G\rangle\otimes|1,1\rangle$ and $|G\rangle\otimes|0,0\rangle$ can be directly obtained. $|G\rangle\otimes|1,-1\rangle$ and $|G\rangle\otimes|1,0\rangle$ belong to the invariant subspace $H_{N/2-2}$ and $H_{N/2-1}$, respectively. One would note that the invariant subspace $H_{N/2-1}$ is two-dimensional. By calculating the evolution state and tracing out the bath, we can obtain the reduced density matrix of the central two-qubit,
\begin{equation}\label{rho2}
\rho_{ab}(t)  =  Tr_{bath}[U(t)\rho_{tot}(0)U^{\dagger}(t)]
  =  \left(\begin{array}{cc}
\widetilde{\mathcal{A}} & \widetilde{\mathcal{Z}}\\
\widetilde{\mathcal{Z}}^{*} & \widetilde{\mathcal{C}}
\end{array}\right)\oplus\left(\begin{array}{cc}
\widetilde{\mathcal{B}} & \widetilde{\mathcal{Y}}\\
\widetilde{\mathcal{Y}} & \widetilde{\mathcal{B}}
\end{array}\right),
\end{equation}

with
\begin{eqnarray}
\widetilde{\mathcal{A}} & = & \frac{1+k_{z}}{4}(1+|\widetilde{U}_{13}^{\prime\prime}|^{2})+\frac{1+k_{x}+k_{y}-k_{z}}{4}|\widetilde{U}_{12}^{\prime}|^{2},\nonumber\\
\widetilde{\mathcal{B}} & = & \frac{1+k_{z}}{8}|\widetilde{U}_{23}^{\prime\prime}|^{2}+\frac{1+k_{x}+k_{y}-k_{z}}{8}|\widetilde{U}_{22}^{\prime}|^{2}\nonumber\\
&  &+\frac{1-k_{x}-k_{y}-k_{z}}{8},\nonumber\\
\widetilde{\mathcal{C}} & = & \frac{1+k_{z}}{4}|\widetilde{U}_{33}^{\prime\prime}|^{2},\nonumber\\
\widetilde{\mathcal{Y}} & = & \frac{1+k_{z}}{8}|\widetilde{U}_{23}^{\prime\prime}|^{2}+\frac{1+k_{x}+k_{y}-k_{z}}{8}|\widetilde{U}_{22}^{\prime}|^{2}\nonumber\\
&  &-\frac{1-k_{x}-k_{y}-k_{z}}{8},\nonumber\\
\widetilde{\mathcal{Z}} & = & \frac{k_{x}-k_{y}}{4}e^{i(N+2)t}\widetilde{U}_{33}^{\prime\prime*},
\end{eqnarray}
where $\widetilde{U}_{ij}^{\prime}(i,j=1,2)$ and $\widetilde{U}_{ij}^{\prime\prime}(i,j=1,2,3)$ denote the matrix elements of the dynamics evolution operator in invariant subspaces $H_{N/2-1}$ and $H_{N/2-2}$.

As shown above, all nonzero elements of reduced density matrices are time dependent. On contrast, when central two-qubit couples to an $XY$ chain, the diagonal elements of reduced density matrix remain constant~\cite{Liu1}. This implies that effects of LMG bath on quantum correlation would be different from that of $XY$ chain.

\section{Quantum correlation of a central two-qubit}\label{QC}
Quantum discord is proposed to quantify all the nonclassical correlation in the bipartite quantum system. For two-qubit state $\rho_{ab}$, it is defined as difference between two versions of mutual information extended from classical to quantum system \cite{Zurek1,Vedral2},
\begin{equation}
Q(\rho_{ab})=I(\rho_{ab})-C(\rho_{ab}).
\end{equation}
Here $I(\rho_{ab})=S(\rho_{a})+S(\rho_{b})-S(\rho_{ab})$ is the
quantum mutual information which measures the total correlation in state $\rho_{ab}$, where $S(\rho)=-Tr(\rho\log_{2}\rho)$ is von Neumann entropy, and $\rho_{a(b)}$ is the reduced density matrix of $\rho_{ab}$. $C(\rho_{ab})$ as a quantifier of classical correlation is the maximum of quantum conditional entropy by performing one side measurement~\cite{Vedral2},
\begin{equation}
C(\rho_{ab})=\max_{\{\Pi_{b}^{(i)}\}}\bigg\{S(\rho_{a})-\sum_{i}p_{i}S(\rho_{a}^{(i)})\bigg\},
\end{equation}
where $\{\Pi^{(i)}_{b}\}$ denotes a set of von Neumann projectors on $b$, $\rho_{a}^{(i)}=Tr_{b}(\Pi_{b}^{(i)}\rho_{ab}\Pi_{b}^{(i)})/p_{i}$ is the state of $a$ after obtaining outcome $i$ on $B$, and $p_{i}= Tr_{ab}(\Pi_{b}^{(i)}\rho_{ab}\Pi_{b}^{(i)})$.

Since the condition $S(\rho_{a})=S(\rho_{b})$ is satisfied in expression~(\ref{rho1}) and~(\ref{rho2}), the measurement of classical correlation assumes unique values, namely, $C(\rho_{ab})$ is irrespective of whether the measurement is performed on qubit $a$ or $b$. We choose the complete set of von Neumann measurement to be of qubit $b$, i.e., $\{\Pi_{b}^{(i)}=|i\rangle\langle i|\}(i=1,2)$, with $|1\rangle=\cos\theta|\uparrow\rangle+e^{i\phi}\sin\theta|\downarrow\rangle$ and $|2\rangle=e^{-i\phi}\sin\theta|\uparrow\rangle-\cos\theta|\downarrow\rangle$. It is straightforward to calculate the quantum discord of  Eq.~(\ref{rho1})~\cite{Fanchini1},
\begin{equation}\label{qd1}
Q(\rho_{ab})=\min\{D_{1},D_{2}\},
\end{equation}
where
\begin{eqnarray}
D_{1} & = & S(\rho_{a})-S(\rho_{ab})-\mathcal{A}\log_{2}\bigg(\frac{\mathcal{A}}{\mathcal{A}+\mathcal{B}}\bigg)-\mathcal{B}\log_{2}\bigg(\frac{\mathcal{B}}{\mathcal{A}+\mathcal{B}}\bigg)\nonumber\\
 &  & - \mathcal{C}\log_{2}\bigg(\frac{\mathcal{C}}{\mathcal{B}+\mathcal{C}}\bigg)-\mathcal{B}\log_{2}\bigg(\frac{\mathcal{B}}{\mathcal{B}+\mathcal{C}}\bigg)
\end{eqnarray}
and
\begin{equation}
D_{2}=S(\rho_{a})-S(\rho_{ab})-\frac{1-\Theta}{2}\log_{2}\frac{1-\Theta}{2}-\frac{1+\Theta}{2}\log_{2}\frac{1+\Theta}{2},
\end{equation}
with $\Theta=\sqrt{(\mathcal{A}-\mathcal{C})^{2}+4(|\mathcal{Y}|+|\mathcal{Z}|)^{2}}$.

In order to compare the dynamics of quantum discord with the dynamics of entanglement, we choose the entanglement of formation as a quantifier of entanglement~\cite{Bennett1}. The entanglement of formation for two-qubit is a monotonically increasing function of concurrence~\cite{Wootters1},
\begin{equation}\label{cu1}
E=-\Lambda\log_{2}\Lambda-(1-\Lambda)\log_{2}(1-\Lambda),
\end{equation}
where $\Lambda=\frac{1}{2}(1+\sqrt{1-\zeta^{2}})$ with the concurrence $\zeta$. For the density matrix as Eq.~(\ref{rho1}), one can directly calculate the concurrence as $\zeta=2\max\{0, \mathcal{Y}-\sqrt{\mathcal{AC}}, |\mathcal{Z}|-\mathcal{B}\}$. By replacing the variables $\mathcal{A}$, $\mathcal{B}$, $\mathcal{C}$, $\mathcal{Y}$, and $\mathcal{Z}$ with $\widetilde{\mathcal{A}}$, $\widetilde{\mathcal{B}}$, $\widetilde{\mathcal{C}}$, $\widetilde{\mathcal{Y}}$ and $\widetilde{\mathcal{Z}}$ in expressions~(\ref{qd1}) and~(\ref{cu1}) respectively, one can directly obtain the quantum discord and entanglement of formation for the density matrix in Eq.~(\ref{rho2}).

\begin{figure}
  \includegraphics[width=6cm]{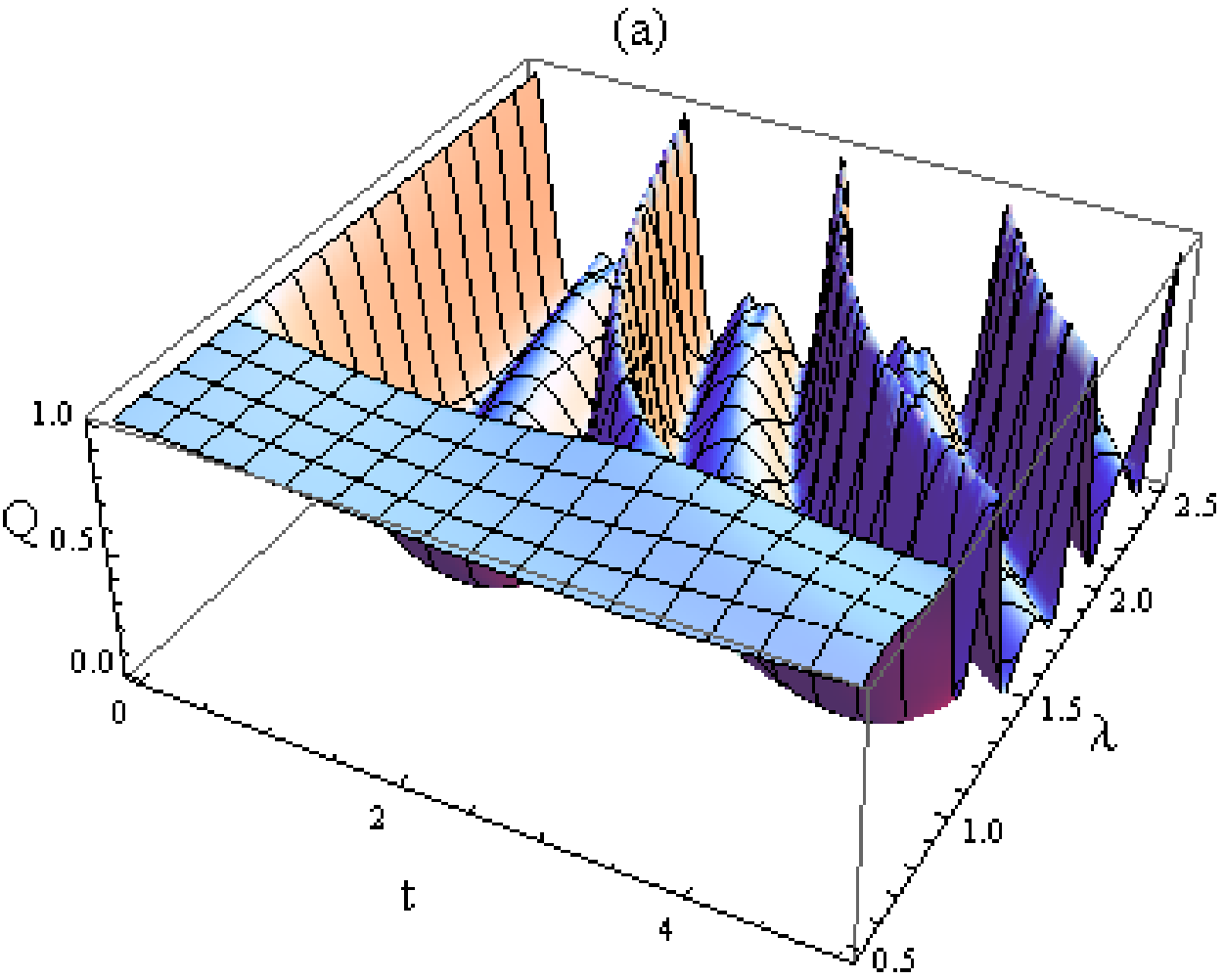}
  \includegraphics[width=6cm]{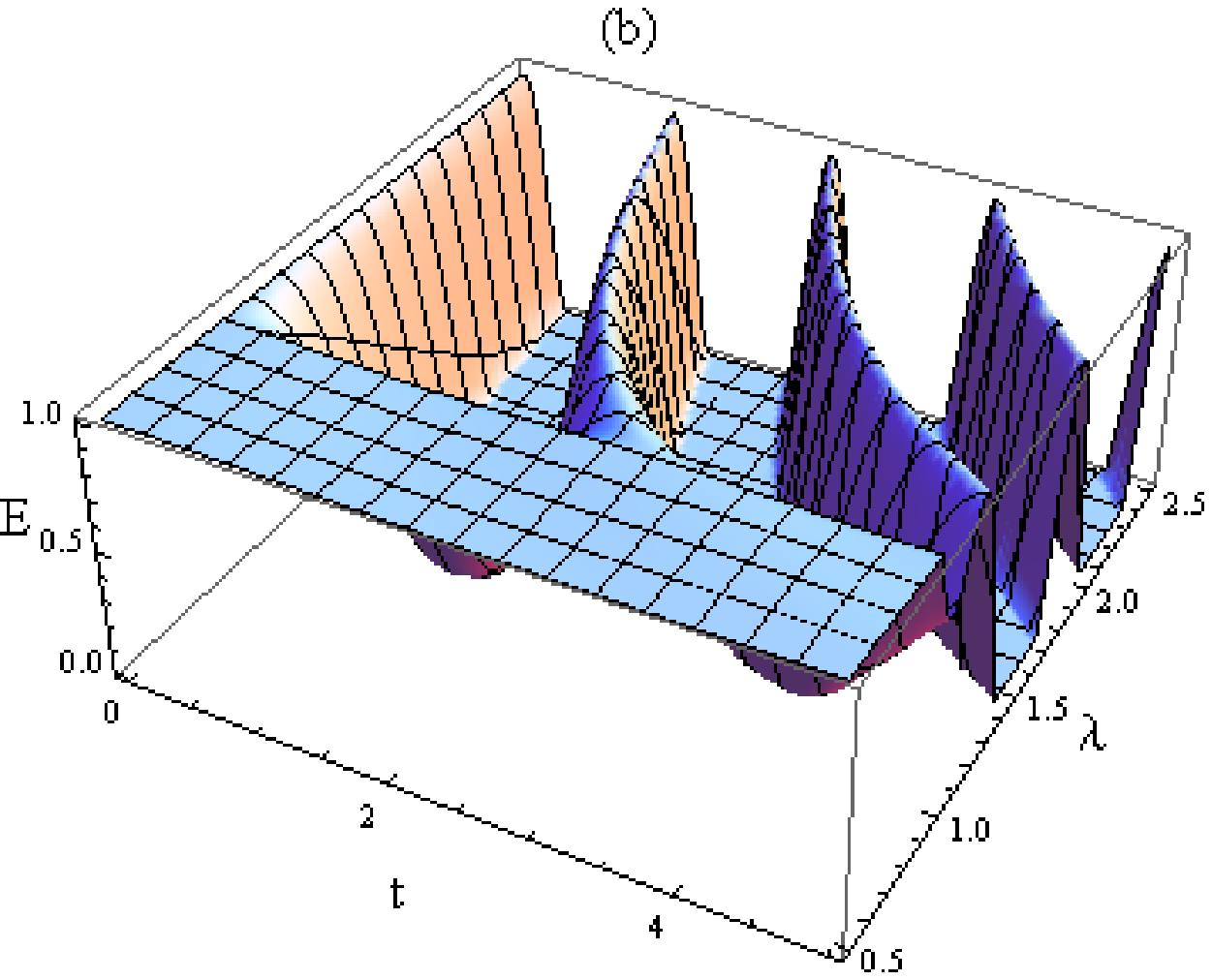}\\
  \caption{(a) quantum discord and (b) entanglement as a function of the coupling strength $\lambda (\lambda^{\prime=\lambda})$ and the time $t$. Other parameters are set as $k_{x}=1,-k_{y}=k_{z}=1$, and $N=500$.}\label{FIG1}
\end{figure}

\begin{figure}
  \includegraphics[width=9cm]{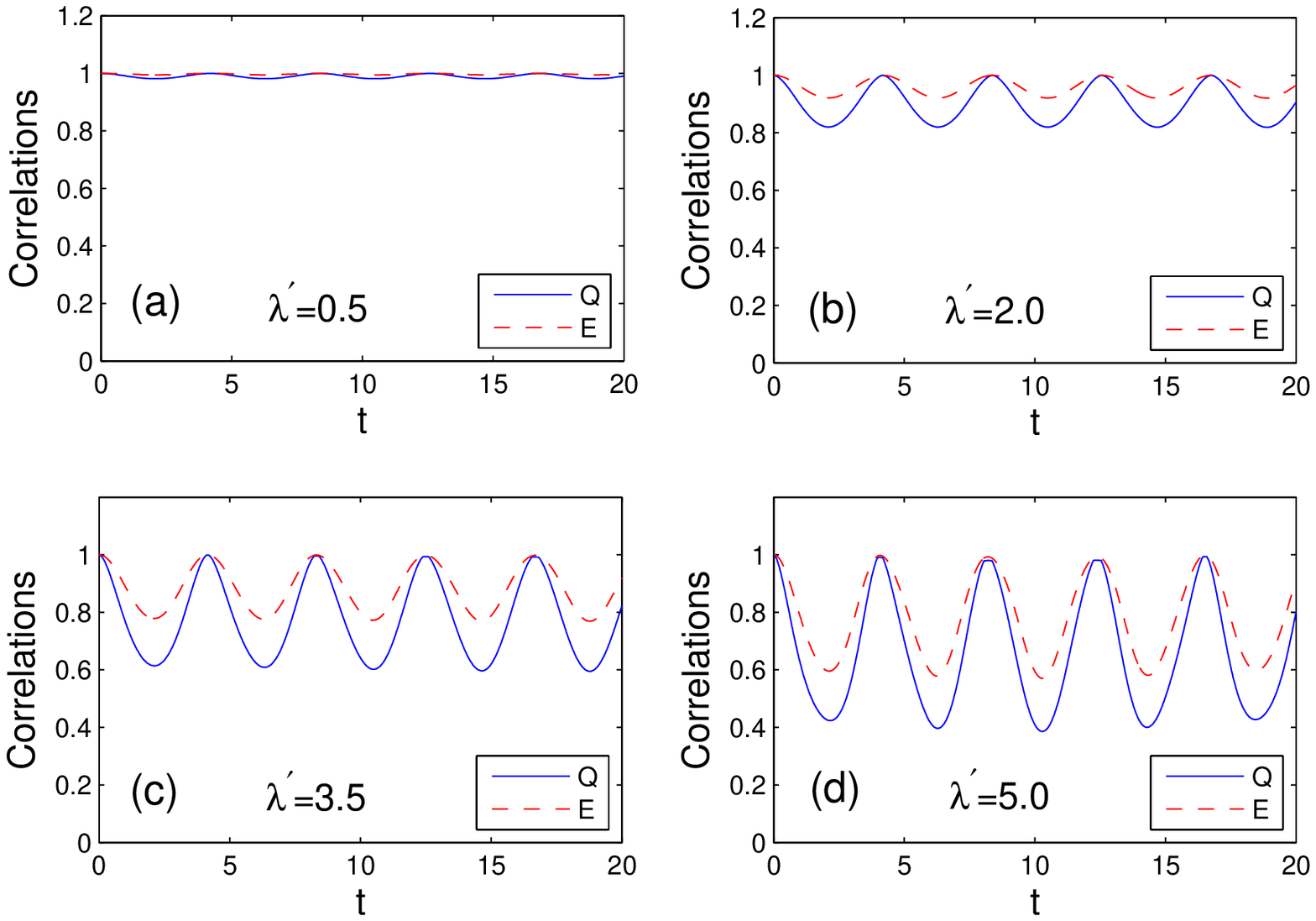}\\
  \caption{Quantum discord (Q) and entanglement (E) as a function of the time $t$ when the bath is in the symmetry-broken phase with $\lambda=0.75$, and (a) $\lambda^{\prime}=0.5$, (b) $\lambda^{\prime}=2.0$, (c) $\lambda^{\prime}=3.5$, (d) $\lambda^{\prime}=5.0$. Other parameters are set as $k_{x}=1,-k_{y}=k_{z}=1$, and $N=1000$.}\label{FIG2}
\end{figure}

\begin{figure}
  \includegraphics[width=9cm]{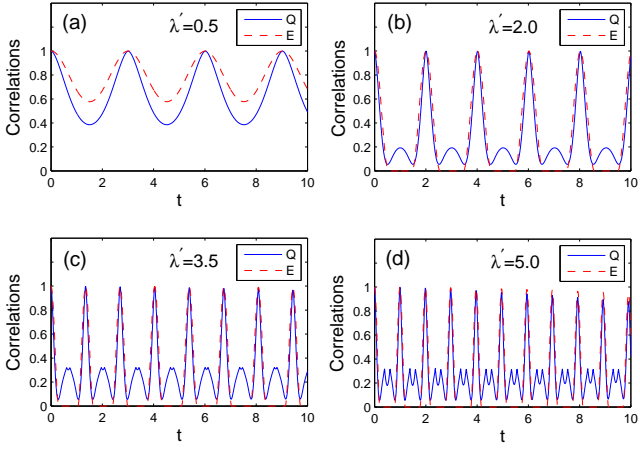}\\
  \caption{Quantum discord (Q) and entanglement (E) as a function of the time $t$ when the bath is in the symmetric phase with $\lambda=1.25$, and (a) $\lambda^{\prime}=0.5$, (b) $\lambda^{\prime}=2.0$, (c) $\lambda^{\prime}=3.5$, (d) $\lambda^{\prime}=5.0$. Other parameters are set as $k_{x}=1,-k_{y}=k_{z}=1$, and $N=1000$.}\label{FIG3}
\end{figure}

Now, let us begin to analyze the dynamics of quantum correlation for the central two-qubit. By setting different values of $k_{\alpha}(\alpha=x,y,z)$, one can easily choose the initial state for the central two-qubit. In this paper, we will consider two cases, namely, the initial state to be pure and mixed. To investigate the dynamics behavior of quantum correlation, it is necessary to know the coupling strength between the cental two-qubit and the bath. First, let us study the case with $\lambda^{\prime}=\lambda$. By setting $k_{x}=-k_{y}=k_{z}=1$, the central two-qubit is assumed to be in the Bell state $\frac{1}{\sqrt{2}}(|\uparrow\uparrow\rangle+|\downarrow\downarrow\rangle)$. In Fig.~\ref{FIG1}, we plot the quantum discord and entanglement as a function of the coupling strength $\lambda$ and the time $t$. It is clear to see that the quantum discord behaves quite differently in each phase of the bath as $\lambda$ goes across $\lambda=1$. Quantum discord remains as a constant unity in the symmetry-broken phase, while it oscillates with time $t$ in the symmetric phase. The similar behavior of entanglement is also observed as depicted in Fig.~\ref{FIG1}(b). Furthermore, Fig.~\ref{FIG1} also shows that entanglement periodically undergoes sudden death while quantum discord has local extrema at the same intervals of time. The common property of the oscillation of quantum correlation is that the frequency can be enhanced by increasing the coupling strength. The phenomena imply that the bath with symmetry-broken phase prefers to preserve the quantum correlation of the central two-qubit in the case  $\lambda^{\prime}=\lambda$. It is clear that the behavior of quantum correlation is quite different from that in $XY$ chain~\cite{Liu1} when the environment approaches the critical point of QPT.

To check the effects of the coupling strength $\lambda^{\prime}$ on the quantum correlation, we do the numerical calculation for the exact expressions (\ref{qd1}) and (\ref{cu1}) for fixed $\lambda$ values. Figure~\ref{FIG2} shows the dynamical behavior of quantum discord and entanglement under different $\lambda^{\prime}$ values when the bath is in the symmetry-broken phase with $\lambda=0.75$. As increasing the values of $\lambda^{\prime}$, the oscillation amplitudes of quantum correlation increase. In Fig.~\ref{FIG3}, we plot the dynamical behavior of quantum discord and entanglement when the bath is in the symmetric phase with $\lambda=1.25$. The effects of the parameter $\lambda^{\prime}$ becomes more significant in the symmetric phase than that in the symmetry-broken phase. The oscillation frequency of the quantum correlation apparently increases as increasing $\lambda^{\prime}$. In addition, it is clearly observed that quantum discord and entanglement synchronously oscillate. Fig.~\ref{FIG2}(a) also shows that the bath with symmetry broken phase preserves the quantum correlation for the case $\lambda^{\prime}<\lambda$. Comparison between Fig~\ref{FIG2} and Fig.~\ref{FIG3} also shows that fluctuations of quantum correlation in the symmetry-broken phase are smaller than that in the symmetric phase, which indicates the bath with symmetry-broken phase preserves a relatively stable quantum correlation.

\begin{figure}
  \includegraphics[width=9cm]{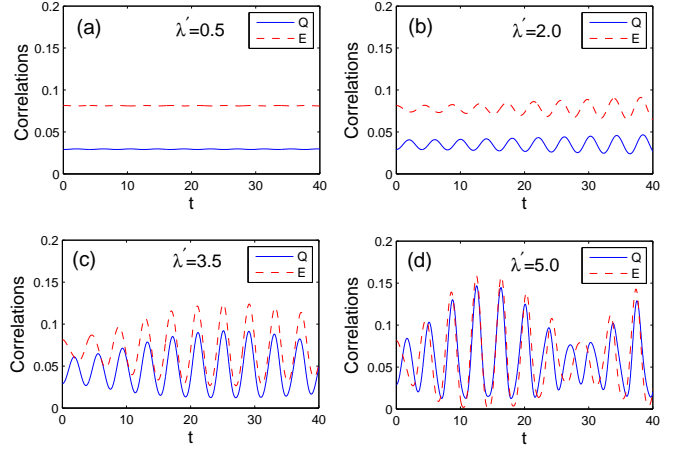}\\
  \caption{Quantum discord (Q) and entanglement (E) as a function of the time $t$ when the bath is in the symmetry-broken phase with $\lambda=0.75$, and (a) $\lambda^{\prime}=0.5$, (b) $\lambda^{\prime}=2.0$, (c) $\lambda^{\prime}=3.5$, (d) $\lambda^{\prime}=5.0$. Other parameters are set as $k_{x}=1,-k_{y}=k_{z}=0.2$, and $N=1000$.}\label{FIG4}
\end{figure}

\begin{figure}
  \includegraphics[width=9cm]{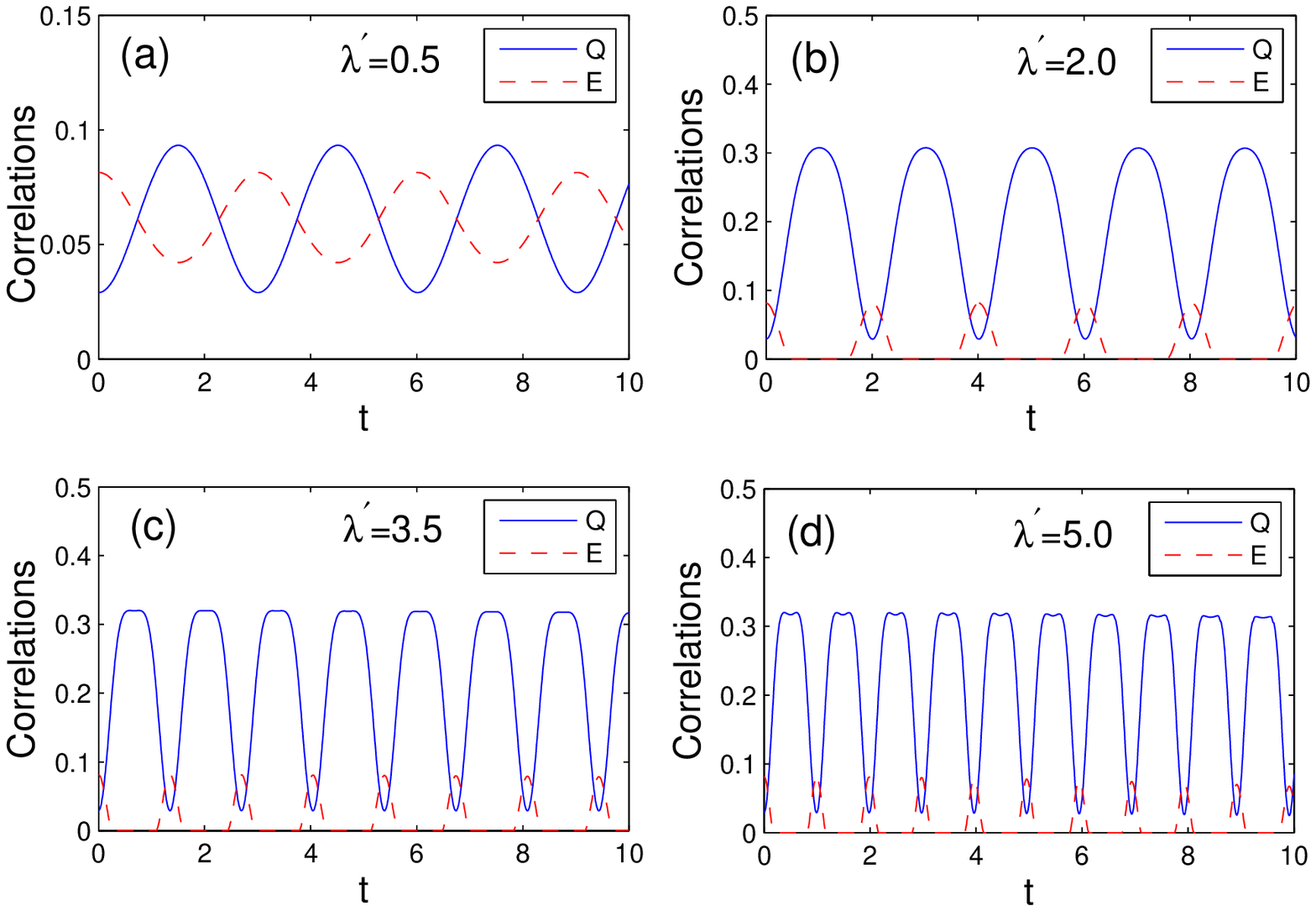}\\
  \caption{Quantum discord (Q) and entanglement (E) as a function of the time $t$ when the bath is in the symmetric phase with $\lambda=1.25$, and (a) $\lambda^{\prime}=0.5$, (b) $\lambda^{\prime}=2.0$, (c) $\lambda^{\prime}=3.5$, (d) $\lambda^{\prime}=5.0$. Other parameters are set as $k_{x}=1,-k_{y}=k_{z}=0.2$, and $N=1000$.}\label{FIG5}
\end{figure}

To study the dynamical properties of quantum discord when the central two-qubit is initially prepared in mixed state, we set the parameters $k_{x}=1, -k_{y}=k_{z}=0.2$, namely, the initial state is Bell-diagonal state with $\rho_{AB}(0)=0.6|\Phi^{+}\rangle\langle\Phi^{+}|+0.4|\Psi^{+}\rangle\langle\Psi^{+}|$, where $|\Phi^{+}\rangle=\frac{1}{2}(|\uparrow\uparrow\rangle+|\downarrow\downarrow\rangle)$ and $|\Psi^{+}\rangle=\frac{1}{2}(|\uparrow\downarrow\rangle+|\downarrow\uparrow\rangle)$ are Bell states. In Fig.~\ref{FIG4}, we plot quantum discord and entanglement as a function of the time $t$ under different $\lambda^{\prime}$ when the bath is in the symmetry-broken phase. As increasing the values of $\lambda^{\prime}$, the oscillation amplitude of quantum correlation also changes with time $t$. Quantum discord and entanglement in this case oscillate in a more complicated way than that of the former cases illustrated in Fig.~\ref{FIG2}. Figure~\ref{FIG5} shows the dynamical behavior of quantum discord and entanglement when the bath is in the symmetric phase. The behavior of quantum correlation is quite different from the case of pure state with the same condition. In the cases of $\lambda^{\prime}=0.5$ and $2.0$, one can find that quantum discord gets maximum while entanglement gets minimum, and vice versa. As increasing the values of $\lambda^{\prime}$, the maximum values of quantum discord is also significantly enhanced while entanglement undergoes periodicity sudden death. This indicates the bath can induce quantum discord while destroyed entanglement at the same condition.

\section{Conclusions}\label{con1}
In summary, we have investigated the dynamics of quantum discord and entanglement for a central two-qubit coupled to an isotropic Lipkin-Meshkov-Glick bath, which has symmetry-broken and symmetric phase. In the bath with symmetry-broken phase, quantum discord and entanglement of central two-qubit can remain stable values with slight fluctuations as tuning the coupling strength between the central two-qubit and the bath being close to the coupling strength of spins of the bath. However, in the bath with symmetric phase, quantum discord and entanglement vary periodically with sharp fluctuations. Furthermore, the fluctuations can be strengthened by increasing the coupling strength between the central two-qubit and the bath, and the oscillation frequency of quantum correlation in the symmetric phase can be enhanced. Nevertheless, quantum correlation oscillate in the symmetry-broken phase with smaller fluctuations than that in the symmetric phase. Our results imply that the bath with symmetry-broken phase prefers to preserve the quantum correlation in this sense. The critical point of quantum phase transition of the bath is clearly revealed by the distinct behaviors of quantum correlation in the two phases. In addition, in the case of central two-qubit with initial pure state, entanglement behaves as similarly as quantum discord except that entanglement occurs sudden death while quantum discord is still nonzero. In the case of central two-qubit with initial mixed state, quantum discord and entanglement become drastic. When entanglement suddenly disappears, quantum discord is significantly enhanced, which imply that quantum discord is more robust than entanglement under decoherence. Our results show that the isotropic LMG bath is able to provide an environment where the central two-qubit remains their coherence, and it may have potential application in quantum information processes. A further study on the effects of anisotropic LMG bath is expected.

\begin{acknowledgments}
This work was partially supported by the NSF of China (Grant No.
11075101).
\end{acknowledgments}

\end{document}